\documentstyle[prl,epsf,aps]{revtex}
\begin{document}
\draft
\twocolumn[\hsize\textwidth\columnwidth\hsize\csname
@twocolumnfalse\endcsname


\title{Variational Study of the Spin-Gap Phase of the One-Dimensional t-J Model}

\author{Y. C. Chen$^{1}$ and T. K. Lee$^{2}$}
\address{
$^{1}$Department of Physics, Tunghai University,
Taichung 400, Taiwan, Republic of China \\
$^{2}$Dept. of Physics, Virginia Tech, Blacksburg, VA 24061}

\date{\today}
\maketitle

\begin{abstract}
We propose a correlated spin-singlet-pairs wave function
to describe the spin-gap phase
of the one-dimensional $t-J$ model at low density. 
Adding a Jastrow factor with a variational parameter, $\nu$,  first introduced 
by Hellberg and Mele\cite{hm1},
is shown to correctly describe the long-range behavior expected for the 
Luther-Emery phase. Using the variational Monte Carlo method we establish a
relation between $\nu$ and the Luttinger exponent $K_\rho$,
 $K_\rho=\frac{1}{2\nu}$. 
\end{abstract}


\vskip2pc]
\narrowtext
\newpage
The high temperature superconductivity(HTSC) has continued to be one of the
important current issues in the field of condensed matter physics.
It is generally believed that the two-dimensional (2D) $t-J$
model is a good working model for the theory of HTSC,
which involves the
basic interactions of the copper-oxygen planes. However, recently the
one-dimensional (1D) version of the model has received much attentions.
Inspired by the unusual normal-state properties of high temperature
superconductors (HTS),
Anderson\cite{pwa} proposed that 
HTS may be described
as the Tomonaga-Luttinger liquid (TLL)\cite{1dtll}
instead of the conventional
Landau's Fermi liquid. Since the 
TLL is well studied in the 1D model, many of its properties
give us useful references in the study of the 2D model. 
For example, the shift of
the characteristic momentum from $2k_F$ to $2k_{F}^{SF}$ of spinless fermions
(SF) in the density-density
correlation function has been used to argue for spin and charge
separation\cite{novel,bill}.
Recently, there are more experimental evidences for a spin-gap phase \cite{sg}
in
HTS. So far there is no good account of this phase from the 2D model.
It turns out that there is also a phase with a spin gap in the 1D model. Recent
studies \cite{newphase,hm2} have identified 
this phase as the Luther-Emery (LE) phase\cite{luther}. A more careful study
of this phase in 1D would help us to gain insights in dealing with the
2D.  In this paper we will present a wave function that catches the
essence of this LE phase. 

The 1D $t-J$ model is
defined as
\begin{eqnarray}
 H_{tJ} &=& -t\sum_{i\sigma} (c^+_{i\sigma} c_{i+1\sigma}+h.c.) \nonumber \\
        &+& J\sum_{i}({\bf S}_i\cdot{\bf S}_{i+1}-{1\over4} n_in_{i+1}),
\end{eqnarray}
with the constraint of no double occupancy. By diagonalizing the 16-site chain,
Ogata et al.\cite{ogata} has found two phases in the phase diagram of the 1D
$t-J$ model.
For very large J/t holes and spins phase separate. With decreasing J 
the system is described as a TLL which exhibits power-law correlations with
exponents charaterized by a single parameter, $K_\rho$. For TLL there is no gap
in both spin and charge excitations. In addition to the momentum
distribution, the important correlation functions such as spin-spin,
density-density and pair-pair, all have power-law decay: 
\begin{eqnarray}
\left<S_z(r)S_z(0)\right> & \sim & A_0 r^{-2} + A_1 cos (2k_Fr) r^{-\lambda_S} 
\nonumber \\
                      & + & A_2 cos (4k_Fr) r^{-\lambda_4} ,
\end{eqnarray}
\begin{eqnarray}
\left<N(r)N(0)\right> & \sim & B_0 r^{-2} + B_1 cos (2k_Fr) r^{-\lambda_N} 
\nonumber \\
                      & + & B_2 cos (4k_Fr) r^{-\lambda_4} ,
\end{eqnarray}
\begin{equation}
P(r)=\left<\Delta^\dagger(r)\Delta(0)\right> \sim C_0 r^{-\lambda_P}
\end{equation}
where
$\Delta(i)=C_{i\uparrow}C_{i+1\downarrow}-C_{i\downarrow}C_{i+1\uparrow}$,
and some of the exponents are:
$\lambda_S=K_\rho+1$, 
$\lambda_N=K_\rho+1$ and $\lambda_P=\frac{1}{K_\rho}+1$.

Using the ground-state projection method, two groups \cite{newphase,hm2}
 have recently found
a third phase, the
Luther-Emery(LE)\cite{luther} phase with a non-zero spin gap in the region of
low electron density and high interaction strength.  
Due to the spin gap, the spin-spin correlation function decays
exponentially with distance. It has
similar power-law correlation functions for density-density and
pair-pair as the TLL but  with different exponents:
$\lambda_N=K_\rho$ and $\lambda_P=\frac{1}{K_\rho}$.

Variational approaches have been very successful in the study of the
phase diagram. In particular, 
Hellberg and Mele (HM)\cite{hm1} have shown that the TLL is very well
represented by  
 a wave function with long-range
Jastrow correlations.
The HM wave function is defined as
\begin{equation}
\left|HM\right>=\Pi_{i>j}\left[\frac{L}{\pi}sin\left(\frac{\pi}{L}
(r_i-r_j)\right)\right]^\nu\Phi_F ,
\end{equation}
where $r_i$ denotes the hole position, $L$ is the total number of sites
and $\Phi_F$ is the projected ideal Fermi gas wave function. The holes repel each other
when $\nu$ is positive and attract otherwise.
HM also showed that with tuning $\nu$ for different $J/t$ and densities
this wave function can reproduce the phase diagram 
of the 1D $t-J$ model. Not only the energy is quite accurate it also
produces the correct power-law correlations that are the signatures of the
TLL. Specifically they have found  
\begin{equation}
K_\rho=\frac{1}{2\nu+1}.
\end{equation}
However 
$\left|HM\right>$
 fails to predict the spin-gap phase \cite{newphase,hm2} which is 
a LE phase instead of a TLL. 

In the study of the spin-gap phase, using
 the exact pairing wave function of two electrons in an
infinite chain as a basis, Chen and
Lee\cite{newphase} (CL) proposed a singlet-pair (SP) wave function,
\begin{equation}
\left|SP\right>= Pd  [\sum_{n=1}^{\infty} h^{n-1}
b_{n}^{+}]^{N_{e}/2} \left|0\right> ,
\end{equation}
where $h=2t/J$. $N_{e}$ is the total number of electrons,
$Pd$ is the projection
operator that forbids two particles occupying the same site and the operator
$b_{n}^{+}= \sum_{i} C_{i\uparrow}^{+} C_{i+n\downarrow}^{+} -
C_{i\downarrow}^{+} C_{i+n\uparrow}^{+}$. Without any tuning parameters CL
showed that $\left|SP\right>$ has lower energy than $|HM\rangle$ and more
significantly it has the correct 
short distance spin-spin correlation which characterizes
the LE phase. 
However with increasing particle density there will be correlations
among pairs and holes. 
And more seriously,
$\left|SP\right>$ is of a particular form of the projected BCS
wave function (or RVB)
hence produces a long-range pairing order, which is inconsistent with
the LE phase with a power-law correlation in pairing. 
Taking $h$ as a tuning parameter will not help to suppress the
long-range order.

In view of the power-law correlation produced by the Jastrow factor in
$\left|HM\right>$, a natural way to modify the wave function
 $\left|SP\right>$ in order 
to correctly represent the pairing correlation in the LE phase is
to add the Jastrow factor to 
 $\left|SP\right>$. This new trial function 
denoted as $\left|CSP\right>$ is
\begin{equation}
\left|CSP\right>=\Pi_{i>j}\left[\frac{L}{\pi}sin\left(\frac{\pi}{L}
(r_i-r_j)\right)\right]^\nu\left|SP\right>.
\end{equation}

Here we present our results of $\left|CSP\right>$ by using the variational Monte
Carlo method. The closed-shell boundary condition
is used for all the data presented in this paper.
The correlation functions either in real space or in
Fourier space are shown in Fig.1 for various $\nu$.
The pairing correlation function is defined in Eq.(4).
The variational parameter is $h=2/3$ for  $J/t \simeq 3$
at the electron density $n_e=1/6$. 
Open circles are the results of $\left|HM\right>$ 
with $\nu=-0.5$
which is the optimal wave function within $\left|HM\right>$ 
in the spin-gap region of 1D $t-J$ model. 
Comparing the variational
energies among several wave functions 
 for $J/t=3$
we find that the lowest energy is obtained by $\left|CSP\right>$
with $\nu=0.1$ (open triangles).
Fig.1(a) shows that 
in this region of $J/t$, 
 $\left|HM\right>$ 
underestimates
the pairing correlation 
 as compared to the 
 $\left|CSP\right>$. 
This 
has been pointed out in previous studies using the ground-state projection
method \cite{newphase,hm2}. 
On the other hand, the long-distance  bahavior of the pairing correlation of
$\left|SP\right>$ ($\nu=0$, filled circle) seems to 
lead to a long-range pair ordering as mentioned above.
With a non-zero $\nu$ the long-range behavior has changed and we will
show below that it has the power-law dependence as expected from
the  LE phase.

\begin{figure}[m]
\epsfysize=15.0cm\epsfbox{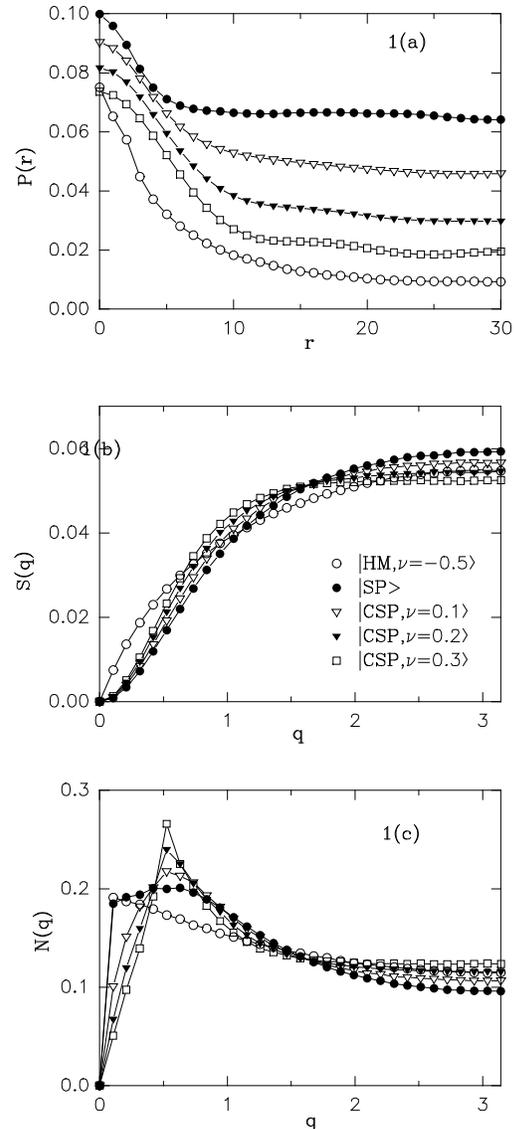}
\caption{
Variational Monte Carlo results of (a)pairing correlations
(shown in real space), (b)spin and (c)density structure factors
of $\left|HM,\nu=-0.5\right>$ and $\left|CSP\right>$ for 
several different $\nu$.
The system has
10 electrons in 60 sites. Symbols 
are defined in Fig.1(b).}
\label{cspfig1}
\end{figure}

    Spin ($S(q)$) and density ($N(q)$) 
structure factors  are plotted in Figs. 1(b) and 1(c) respectively, with 
the same parameters as in Fig.1(a). 
It is easy to show that $S(q)$ should be quadratic at small $q$ if there is
a gap in spin excitations \cite{hohenberg}.
This indeed has been found in Fig. 1(b) for
 $\left|SP\right>$ 
 and $\left|CSP\right>$. 
More interestingly, this behavior is quite robust, it hardly 
changes with the variation of the
exponent $\nu$ in contrast to the dramastic changes occurred in pairing 
and density correlations. The reason, we believe, is that the wave functions
 $\left|SP\right>$ 
 and $\left|CSP\right>$ 
 only have
 short-range pairs.  This
  preserves the spin gap even 
in the presence of strong density fluctuations induced by the Jastrow
correlation factor.
\begin{figure}[m]
\epsfysize=15.0cm\epsfbox{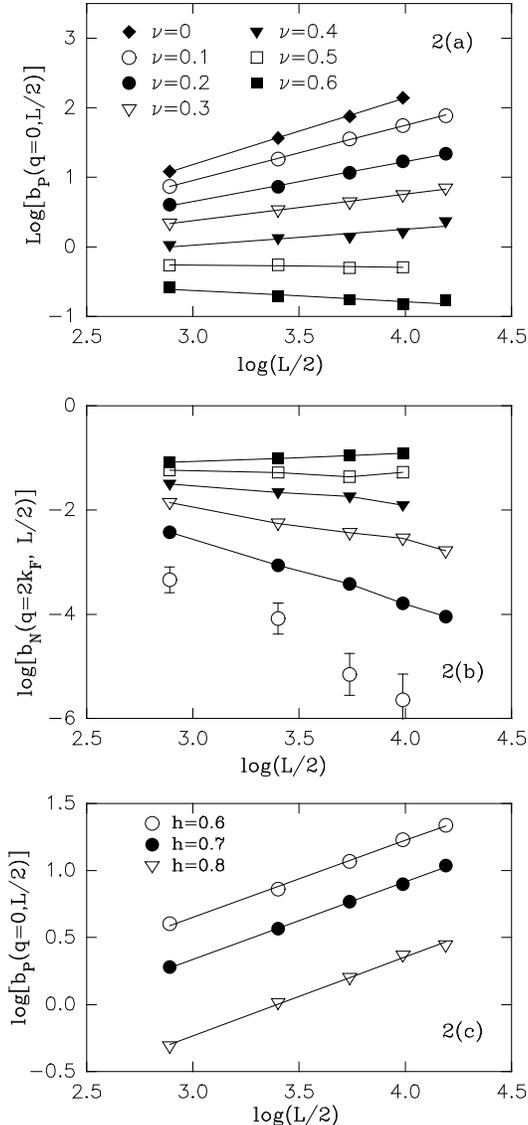}
\caption{
Finite-size scaling of the quantities (a)$b_P(L/2)$ and (b)$b_N(L/2)$
as defined in Eqs.(9) and (10) for $\left|CSP\right>$ with a fixed $h=0.6$. 
The electronic density is
fixed at $n_e=\frac{1}{6}$ for lattice sizes ranging from $L=$36 to 132. 
The solid lines are the least-square fits of the data
points. The slops of the lines in (a) and (b) give $1-\lambda_P$ and 
$1-\lambda_N$, respectively. (c) is similar to (a) but for a fixed $\nu=0.2$
and various $h$.
}
\label{cspfig2}
\end{figure}

As shown in Fig.1(c) $\left|HM,\nu=-0.5\right>$
has the maximum of $N(q)$ at $2\pi/L$.
This indicates the system is close to the phase separation.
 However for the
LE phase
one would expect a sharp peak at $2k_F$ and even a divergent one if
$K_\rho<1$. 
For $\left|SP\right>$ the maximum is indeed at $2k_F$ but it is quite broad.  
 As for $\left|CSP\right>$ we find that the peak 
becomes sharper and sharper
as $\nu$  increases. The size of the peak also
grows with $\nu$. By using finite-size scaling we find that the peak diverges
with the lattice size for $\nu>0.5$ (see below). 


In addition to establish the power-law behavior for density-density and 
pair-pair
correlations
we must also find 
a relation between their exponents. 
In order to extract the exponents of the power-law correlations
we consider 
the finite-size scaling behavior of the following quantities \cite{assaad},
\begin{equation}
b_P(L/2)=P(q=0)-P(q=2\pi/L)
\end{equation}
and
\begin{eqnarray}
b_N(L/2)& = & 2N(q=2k_F)-N(q=2k_F+2\pi/L) \nonumber \\
        & - & N(q=2k_F-2\pi/L),
\end{eqnarray}
where $P(q)$ is the Fourier transform of $P(r)$.
In Fig.2(a) we plot
$\log\left[b_P(q=0,L/2)\right]$ versus $\log(L/2)$
to obtain the exponent
$1-\lambda_P$ as the slope of the linear fit of the data. 
A similar analysis of $b_N(q=2k_F,L/2)$ gives the exponent $1-\lambda_N$.
We have used lattice sizes ranging from $L$=36 to 132.
The very well linear fit of these plots supports our conclusion of
power-law behavior in pairing and density correlations. The exponents obtained
in these plots are tabulated in Table I. We find that the relation
$\lambda_P=\frac{1}{\lambda_N}$ is satisfied with the data in Table I, i.e.,
the scaling relation expected for the
 LE phase is recovered\cite{note}. Additionally we
find that within the error bars  
the variational parameter $\nu$ is related to the critical exponent
$K_\rho$ of the LE phase in a simple 
relation \cite{hellberg}
\begin{equation}
K_\rho=\frac{1}{2\nu}.
\end{equation} 
Correspondingly we have $\lambda_P=\frac{1}{2\nu}$ and $\lambda_N=2\nu$.

\vspace{0.5cm}
{\bf TABLE I.} Critical exponents $\lambda_P$($\lambda_N$) of 
pairing (density) correlations for $\left|CSP\right>$
with various varational parameters $\nu$. The numbers in the parentheses
show the error in the last-digit. 
The last row is the predicted
values of $K_\rho$ by Eq.(11).
\begin{center}
\begin{tabular}{c | c c c c c c c c} \hline
\hline
 $\nu$       & 0.1 & 0.2 & 0.3 & 0.4 & 0.5 & 0.6 \\
\hline
 $\lambda_P$ &0.21(2)&0.43(3)& .62(2)&0.77(8)&1.03(4)&1.16(8) \\
\hline
 $\lambda_N$ &     & 2.3(4) & 1.63(8) & 1.32(8) & 1.06(8) & 0.84(5) \\
\hline
 $K_\rho$ & 5 & 2.5 & 1.667 & 1.25 & 1.0 & 0.833 \\
\hline
\end{tabular}
\end{center}
\vspace{0.5cm}

The finite-size scaling of pairing correlations
for a given
$\nu$(=0.2) and various $h$
is shown in Fig.2(c).
 We find that all the data fit to lines of the same 
slope hence of the same exponent. Therefore we conclude that the long-range
power-law correlations are controlled by the Jastrow factors, irrespective 
of the parameter $h$ which controlls the short-range properties.

\begin{figure}[m]
\epsfysize=9.00cm\epsfbox{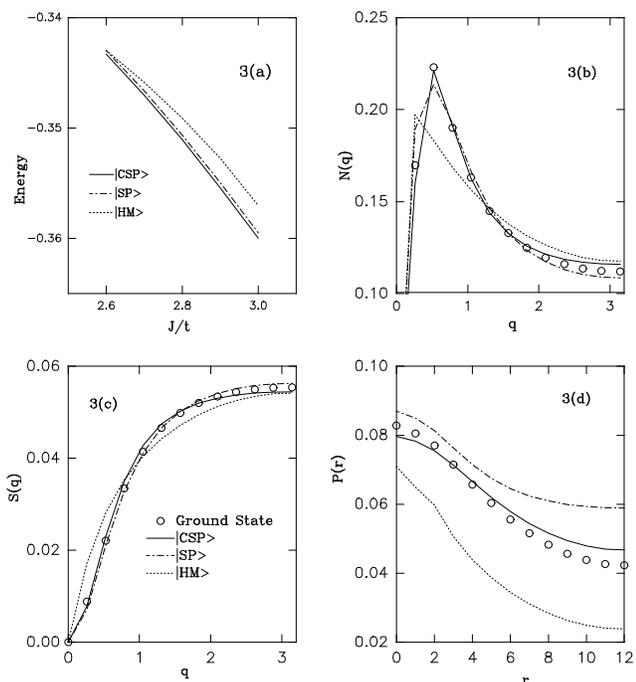}
\caption{
(a)Variational energies, 
(b)Density and (c) spin  structure factor, as well as (d) pairing
correlation of the $t-J$ model for 
$\left|CSP\right>$, $\left|SP\right>$ and
$\left|HM\right>$  at the density $n_e=1/6$.
The variational parameters are optimized for $J/t=2.8$.
We take $h=2t/J$ for both $\left|CSP\right>$ and 
$\left|SP\right>$. And $\nu=0.08, -0.5$ for $\left|CSP\right>$ and
$\left|HM\right>$, respectively. 
 The lines and symbols are labled 
 in (c).
This system with 4 electrons
in a 24-site chain is also exactly diagonalized by the Lanczos method
and these exact results are shown as the circles.
}
\label{cspfig3}
\end{figure}

   Having established $\left|CSP\right>$ as a good wave function to describe
the LE phase, now we like 
to investigate 
if it is also a proper
wave function to faithfully represent the ground state. 
The 
varational energies of $\left|HM\right>$, $\left|SP\right>$ and 
$\left|CSP\right>$ at the density $n_e=1/6$ 
are shown in Fig. 3(a).
We observe that $\left|CSP\right>$
and $\left|SP\right>$ are very close in energy for large  $J/t$. 
By using the power method 
\cite{newphase} we find that the variational energy of $\left|CSP\right>$ is
about $0.3\%$ above the
ground state energy. Although $\left|CSP\right>$ and $\left|SP\right>$ have 
similar energies, they have significant differences in their 
long-range correlations as indicated above.
In Fig.3(b) - (d) we show the 
density and spin structure factors 
and pairing correlations for $\left|HM\right>$, $\left|SP\right>$ and
$\left|CSP\right>$ optimized for $t-J$ model with $J/t=2.8$. 
This system contains 
4 electrons in a 24-site chain. For this system the exact ground state can be 
obtained by the Lanczos diagonalization method. 
Results for the ground state are 
shown  as the open circles in Fig. 3. We find that for spin correlation
$\left|CSP\right>$ and $\left|SP\right>$ 
have similar quadratic behavior at small $q$ and agree with the exact result
very well. However for density and pairing correlations $\left|CSP\right>$
describes much better than $\left|SP\right>$. The excellent consistency between
the correlations of the ground state and $\left|CSP\right>$ supports the 
expectation of the LE phase in the low density region of the 1D $t-J$ model.

In conclusion, we have presented a wave function to describe the LE
phase in 1D. This wave function  
 $\left|CSP\right>$ 
has
 correlated spin-singlet pairs. This is established by
 using the variational Monte Carlo method. The wave function
  shows exponential dependence for the 
spin-spin correlation and power-law behavior in density-desity and
pair-pair correlations. By finite-size scaling we established the
relation of the varational parameter $\nu$ to the exponent 
$K_\rho$, $K_\rho=\frac{1}{2\nu}$. Comparing with the exact ground state of a
small lattice  we showed that $\left|CSP\right>$ describes
very well the ground state properties of the spin-gap phase of the 1D 
$t-J$ model.

We are grateful to C. S. Hellberg for valuable discussions. 
This work is supported 
by National Science Council under Grant No. NSC85-2112-M-029-005. 
Part of the research was conducted using the resources of the Cornell Theory
Center and  
National Center for High-performance
Computing(NCHC)in Taiwan. We thank their supports.


\begin{references}
\bibitem{note} For $\nu=0.1$, the peak in the density  structure factor
is not sharp enough (Fig.1(c)) and it produces large error bars.
\end{references}
\end{document}